**Radiometric $^{81}$Kr dating identifies 120,000 year old ice at Taylor Glacier, Antarctica**

Christo Buizert[a]*, Daniel Baggenstos[b], Wei Jiang[c], Roland Purtschert[d], Vasilii V. Petrenko[e], Zheng-Tian Lu[c,f], Peter Müller[c], Tanner Kuhl[g], James Lee[a], Jeffrey P. Severinghaus[b] and Edward J. Brook[a]

[a]College of Earth, Ocean and Atmospheric Sciences, Oregon State University, Corvallis OR 97331, USA

[b]Scripps Institution of Oceanography, University of California, San Diego, CA 92093, USA

[c]Physics Division, Argonne National Laboratory, Argonne, IL 60439, USA

[d]Climate and Environmental Physics, University of Bern, CH-3012 Bern, Switzerland

[e]Department of Earth and Environmental Sciences, University of Rochester, Rochester NY 14627, USA

[f]Department of Physics and Enrico Fermi Institute, University of Chicago, Chicago, IL 60637, USA

[g]Ice Drilling Design and Operations, University of Wisconsin, Madison WI 53706, USA

*Corresponding author:

buizertc@science.oregonstate.edu

104 CEOAS Admin Bldg., Oregon State University, Corvallis OR 97333, USA

**Abstract**

We present the first successful $^{81}$Kr-Kr radiometric dating of ancient polar ice. Krypton was extracted from the air bubbles in four ~350 kg polar ice samples from Taylor Glacier in the McMurdo Dry Valleys, Antarctica, and dated using Atom Trap Trace Analysis (ATTA). The $^{81}$Kr radiometric ages agree with independent age estimates obtained from stratigraphic dating techniques with a mean absolute age offset of 6 ± 2.5 ka. Our experimental methods and sampling strategy are validated by 1) $^{85}$Kr and $^{39}$Ar analyses that show the samples to be free of modern air contamination, and 2) air content measurements that show the ice did not experience gas loss. We estimate the error in the $^{81}$Kr ages due to past geomagnetic variability to be below 3 ka. We show that ice from the previous interglacial period (MIS 5e, 130-115 ka before present) can be found in abundance near the surface of Taylor Glacier. Our study paves the way for reliable radiometric dating of ancient ice in blue ice areas and margin sites where large samples are available, greatly enhancing their scientific value as archives of old ice and meteorites. At present, ATTA $^{81}$Kr analysis requires a 40-80 kg ice sample; as sample requirements continue to decrease $^{81}$Kr dating of ice cores is a future possibility.

**Significance statement**

Past variations in Earth's climate and atmospheric composition are recorded in accumulating polar meteoric ice and the air trapped within it. Ice outcrops provide accessible archives of old ice, but are difficult to date reliably. Here we demonstrate $^{81}$Kr radiometric dating of ice, allowing accurate dating of up to 1.5 million year old ice. The technique successfully identifies valuable ice from the previous interglacial period at Taylor Glacier, Antarctica. Our method will enhance the scientific value of outcropping sites as archives of old ice needed for paleoclimatic reconstructions, and can aid efforts to extend the ice core record further back in time.

## Introduction

Ice cores from the Greenland and Antarctic ice sheets provide highly resolved, well-dated climate records of past polar temperatures, atmospheric composition and aerosol loading up to 800 ka before present (1-3). In addition to deep ice cores, old ice can also be found at ice margin sites and blue ice areas (BIAs) where it is exposed due to local ice dynamics and ablation (4-6). Antarctic BIAs have attracted much attention for their high concentration of meteorites, which accumulate at the surface over time (7). More recently BIAs have also been used for paleoclimate studies, as large quantities of old ice are available at the surface where it can be sampled with relative ease (8, 9). Because the ice stratigraphy is exposed laterally along the BIA surface, such ice records are often referred to as horizontal ice cores.

Determining the age of the ablating ice is the main difficulty in using BIAs for climate reconstructions (4). The most reliable method is stratigraphic matching, where dust, atmospheric composition or water stable isotopes of the horizontal core are compared to well-dated, regular ice core records to construct a chronology (10, 11). This technique, however, requires extensive sampling along the ice surface and relatively undisturbed stratigraphy, cannot be used past 800 ka BP (the current limit of the ice core record) and can be ambiguous during some time intervals. Several radiometric methods have been applied to ice dating, all of which have distinct limitations. Radiocarbon dating of trapped $CO_2$ suffers from in situ cosmogenic $^{14}$C production in the ice, e.g. (12). Other methods rely on the incidental inclusion of datable material, such as sufficiently thick Tephra layers (13) or meteorites (7). The terrestrial age of meteorites is not likely to be representative of the surrounding ice, though, because they accumulate near the BIA surface as the ice ablates. A promising new technique uses the accumulated recoil $^{234}$U in the ice matrix from $^{238}$U decay in dust grains as an age marker (14); currently the method still has a fairly large age uncertainty (16-300 ka). Flow modeling can provide useful constraints on the age of BIAs (15-18), but large errors are introduced by the limited availability of field data and necessary assumptions about past flow and ice thickness. Finally, the continued degassing of $^{40}$Ar from the solid Earth allows dating of the ancient air trapped in the ice by evaluating the $\delta^{40}$Ar/$^{38}$Ar and $\delta^{40}$Ar/$^{36}$Ar stable isotope ratios (19), but the current uncertainty in this dating method is also large (180 ka or 11% relative age, whichever is greater).

There is significant scientific interest in obtaining glacial ice dating beyond 800 ka, as such an archive would extend the ice core record further back in time, providing valuable constraints on the evolution of past climate, atmospheric composition, and the Antarctic ice sheet (20). Of particular interest is the middle Pleistocene transition (1200-800 ka BP) and the role of $CO_2$ forcing therein (21, 22). Such old ice can potentially be found in Antarctic BIAs such as the Allan Hills site (23), providing a strong impetus to developing reliable (absolute) dating tools for glacial ice.

Krypton is a noble gas present in the atmosphere at a mixing ratio of around 1 ppm (24), and has two long-lived radioisotopes of interest to earth sciences (25, 26): $^{81}$Kr ($t_{1/2}$ = 2.29 · $10^5$ year) and $^{85}$Kr ($t_{1/2}$ = 10.76 year). $^{85}$Kr is produced in nuclear fission, and released into the atmosphere primarily by nuclear fuel reprocessing plants. $^{81}$Kr is naturally produced in the upper atmosphere by cosmic ray interactions with the stable isotopes of Kr, primarily through spallation and thermal neutron capture (27). The long half-life of $^{81}$Kr allows for radiometric dating in the 50 – 1500 ka age range (28), well past the reach of radiocarbon dating. $^{81}$Kr-Kr dating has already been used to determine the residence time of groundwater in old aquifers (29-31) , and for several reasons has great potential for applications in dating polar ice as well. First, Krypton is not chemically reactive. Second, due to its long residence time $^{81}$Kr is well mixed in the atmosphere. Third, the method does not rely on sporadically occurring tephra, meteoric or organic inclusions in the ice, but is widely applicable as all glacial ice contains trapped air. Fourth, it does not require a continuous or undisturbed ice stratigraphy. Finally, in contrast to $^{14}$C it does not suffer from in situ cosmogenic production in the ice (12). What has precluded its use in ice core science to date is the large sample requirement owing to the small $^{81}$Kr abundance ($^{81}$Kr/Kr = 5 · $10^{-13}$). Recent technological advances in Atom Trace Trap Analysis (ATTA (32-34)) have reduced sample requirements to 40-80 kg of ice, which can realistically be obtained from BIAs and ice margins.

An earlier attempt at $^{81}$Kr dating of ice gave inconclusive results owing to ~50% gas loss and substantial, but poorly quantified (≥20%) contamination with modern air (35). The single sample in that study was extracted by chainsaw from the surface of the Allan Hills BIA site in Antarctica, and the sample quality was compromised by "extensive fracturing of the highly strained ice". In addition, a reliable method for $^{81}$Kr/Kr analysis was lacking at the time. Here we describe the first successful $^{81}$Kr radiometric dating of polar ice using air extracted from four ice samples from the Taylor Glacier blue ice area in Antarctica. The ice at this site is heavily fractured to a depth of ~3 m with sporadic cracks extending to 5 m, and consequently we sampled ice in the 5-15m depth range. Using $^{85}$Kr we demonstrate that our samples are uncontaminated by modern air. We independently date our samples using stratigraphic matching techniques, and show an excellent agreement with the $^{81}$Kr radiometric ages. Our study shows that ice from the previous interglacial period (MIS 5e, 130-115 ka BP), can be found in abundance near the surface of the Taylor Glacier BIA.

**Site description and sample selection**

Taylor Glacier is an outlet glacier of the East Antarctic ice sheet that terminates in the McMurdo Dry Valleys (Figure 1a, 1b) (36). Along the lower ~70km of the glacier the surface mass balance is dominated by sublimation (37), causing outcropping of old ice. Stratigraphic matching of water stable isotopes to the nearby Taylor Dome ice core (38) identified ice in the 11.5-65 ka age range at the glacier surface (10). Here we present new and extensive dating of the near-surface ice using a combination of $CH_4$ and $\delta^{18}O_{atm}$ (i.e. the isotopic composition of atmospheric $O_2$ trapped in the ice). Three separate transects on the glacier were stratigraphically dated; two of them are perpendicular to ice flow (the Main and Downstream transects, marked as M-M' and D-D' in figure 1b), and one of them is parallel to ice flow (the Along-flow profile, marked A-A' in figure 1b).

To test the feasibility and accuracy of $^{81}Kr$ radiometric dating of ice, we obtained four ~350 kg samples from Taylor Glacier. All ice sampling was below 5m depth to avoid gas loss and exchange due to near-surface fractures. We selected sampling locations where the age of the ice is well constrained by matching $CH_4$ mixing ratios and/or $\delta^{18}O_{atm}$ to (Antarctic) deep ice core records; we shall refer to this age estimate as the stratigraphic age of the samples. Because $CH_4$ and $\delta^{18}O_{atm}$ are both atmospheric signals, this method only constrains the age of the trapped gas, rather than the age of the meteoric ice itself. There are three main contributions to the stratigraphic age uncertainty; for our samples we will list the root-sum-square of these. First, there is some ambiguity in linking Taylor Glacier samples to ice core records due to analytical uncertainties and the possible non-uniqueness of the synchronization. Second, the ice core chronologies themselves are subject to uncertainties. For the last 60 ka an annual layer-counted age scale is available for Greenland, to which Antarctic records can be tied using globally well-mixed $CH_4$; beyond this age, ice flow modeling is commonly used to reconstruct the chronology (39-41). The uncertainty in the ice core chronologies can be evaluated by comparing them to independently dated speleothem records showing concomitant events (41-44). Third, the Kr samples contain a spread in ages due to their finite size. We estimate this last effect is only important for the oldest sample where the layers are very strongly compressed.

The first sample (Kr-1) was obtained along the Main transect. The sample is from the Younger Dryas period, which is clearly identified by its characteristic $CH_4$ sequence. Figure 2a shows $CH_4$ measurements along the profile (white dots), which are plotted on top of a well-dated $CH_4$ record from the Antarctic EDML core (45). The top axis shows the distance along the transect in meters; note that the position-age relationship is non-linear. We assign a stratigraphic age of 12.2 ± 0.5 ka to this sample (black dot with error bar), with 0.5 ka synchronization uncertainty and a 0.1 ka ice core age uncertainty (46). The next two samples are located on the along-flow profile; these samples are most clearly identified by their $\delta^{18}O_{atm}$ (figure 2b). The upstream samples of this profile are characterized by relatively depleted $\delta^{18}O_{atm}$ values, which allows to date them unambiguously as early Holocene. Going down glacier the ice gets progressively older; ice with ages between 10-55 ka is found in stratigraphic order 0-15 km downstream of the first measurement of the profile. Sample Kr-2 is located in this segment, and has an age 49 ± 3.6 ka. The Kr-2 synchronization uncertainty is set equal to the ~3 ka sampling resolution of the along-flow profile stratigraphic dating, and the ice core age uncertainty is around 2 ka (43). Past 55 ka the age interpretation is more ambiguous, and ice from Marine Isotope Stage 4 (MIS 4) appears to be absent from the sampling profile. We selected a spot where $\delta^{18}O_{atm}$ measurements are identified with MIS 5a. The age at the location of sample Kr-4 is estimated at 79 ± 4.5 ka, with a synchronization uncertainty set to half the width of the MIS 5a $\delta^{18}O_{atm}$ plateau (4 ka), and a 2 ka ice core age uncertainty. For locations > 16.5 km in the along-flow profile we find unexpectedly enriched $\delta^{18}O_{atm}$ values in the profile, perhaps indicative of MIS 4 ice that has traveled further downstream due to strong folding of the glacial ice. Last, we present $CH_4$ and $\delta^{18}O_{atm}$ data from a second transect further downstream (Downstream transect, Fig. 2c), where the oldest sample (Kr-3) was obtained. The strongly depleted $\delta^{18}O_{atm}$ value of sample Kr-3 (-0.39 ‰) is an unambiguous fingerprint of interglacial origin, which is confirmed by interglacial $CH_4$ levels. The combined $\delta^{18}O_{atm}$ and $CH_4$ measured along the Downstream transect are consistent with ice from MIS 5e. We assign a stratigraphic age of 123.5 ± 4.2 ka to sample Kr-3, with a synchronization error of 3 ka , an ice core age uncertainty of <1 ka (44) and a uncertainty due to the sample age spread of 3 ka. It must be noted that the ice stratigraphy in this lower part of the glacier is strongly disturbed by ice flow, and the sequence shown in Fig 2c is surrounded by folded ice that cannot be dated reliably using $CH_4$ and $\delta^{18}O_{atm}$. With the stratigraphic technique alone we cannot firmly exclude the possibility that the ice originates from the mid-Holocene, which has equally depleted $\delta^{18}O_{atm}$ values. Apart from the four ice samples we took an additional atmospheric sample upwind from the field camp, which was processed identically to the air samples extracted from the ice.

## Results and discussion

The results from our analyses are given in Table 1. $^{81}$Kr measurements are normalized to modern atmospheric Kr, and reported in percent of modern Kr: pMKr = $^{81}R_{SA}/^{81}R_{ST}\cdot 100\%$, with $^{81}R_{SA}$ and $^{81}R_{ST}$ the [$^{81}$Kr/Kr] ratio in the sample and the reference standard (i.e. modern atmosphere), respectively. $^{81}$Kr radiometric ages are calculated from the averaged replicates using $t_{Kr}$ (ka) = $-229\cdot\log_2(^{81}R_{SA}/^{81}R_{ST})$. $^{81}$Kr age uncertainties are calculated using standard error propagation techniques, and include analytical uncertainties (on average 6 pMKr) and the uncertainty in the $^{81}$Kr half-life ($t_{1/2}$ = 229 ± 11 ka). The ATTA analysis measures the concentration of $^{81}$Kr relative to that of the stable isotope $^{83}$Kr. In this work, the isotope ratio has a statistical uncertainty of 3.5% due to $^{81}$Kr counts, and a global systematic uncertainty of 5% in the measurements of the $^{83}$Kr concentration. ATTA analyses in the future will benefit from a newly demonstrated technique that has reduced the systematic uncertainty in the $^{83}$Kr measurement down to 0.6% (47). Figure 1c shows a comparison of the $^{81}$Kr radiometric ages to the independently derived stratigraphic ages, with the dashed line giving the one-to-one slope. Note that in comparing these ages one does not need to consider the ice age-gas age offset ($\Delta$age) in glacial ice, because both the $^{81}$Kr and stratigraphic dating are done on the gas phase. For all four ice samples we find that both ages agree within the analytical uncertainty. On average, the absolute age offset between the dating methods is 6 ± 2.5 ka, which is about a third of the estimated uncertainty in the $^{81}$Kr radiometric ages (dominated by the ATTA analytical uncertainty). The $t_{Kr}$ we obtain for sample Kr-3 (120 ka) clearly identifies this ice as originating from the MIS 5e interglacial period, eliminating any remaining age ambiguity in the stratigraphic dating.

Our analyses show that the integrity of our samples has not been compromised. First, the air content we obtain for our samples is 104 to 111 ±5 mL kg$^{-1}$ after correcting for gas dissolution during the melt-extraction of gases from the ice (Table 1). Measured air content in the nearby Taylor Dome ice core is 99 ± 4 mL kg$^{-1}$ for 85-63 ka BP (Supporting information). Taylor Glacier ice originates on the slopes of Taylor Dome and is expected to have slightly higher air content because of the lower elevation of the deposition site. The air content in our samples is consistent with a deposition elevation between the upper reaches of Taylor Glacier (~1500 m.a.s.l. (36)) and the modern Dome (2365 m.a.s.l. (38)). Second, for all ice samples the $^{85}$Kr activity ($t_{1/2}$ = 10.76 years) is below the ATTA detection limit; values given in Table 1 represent the 90% confidence level upper bound activity (decay corrected). By comparing to our atmospheric sample we estimate a 1.5% upper bound on modern contamination in the ice-extracted Kr. We further analyzed the $^{39}$Ar ($t_{1/2}$ = 269 years) activity of the samples using radioactive decay counting. It must be noted that the sample size is too small for precise $^{39}$Ar analysis. With the exception of sample Kr-1, the $^{39}$Ar activity of the samples is below the detection limit (Table 1). The combination of a negligible $^{85}$Kr activity and measureable $^{39}$Ar activity in sample Kr-1 is puzzling. It can conceivably be due to a contamination event (> 15% of air content) that occurred several decades prior to sampling by an unusually deep fracture in the ice; this interpretation is however contradicted by the $CH_4$ mixing ratio of sample Kr-1, which agrees well with the ice core record (Fig. 2a). Another possibility is a modern contamination of the Ar sample fraction after Ar-Kr separation in the laboratory.

Table 1 furthermore gives the $\delta^{86}$Kr = $\delta^{86}$Kr/$^{82}$Kr stable Kr isotope ratio (divided by 4 to show fractionation per unit mass difference) measured on the sample fraction that remained after replicate ATTA $^{81}$Kr/$^{83}$Kr analysis ($\delta^{86}$Kr analysis on sample Kr-3 was compromised by a system leak). For all samples we observe a 2.4-3.5 ‰ enrichment in $\delta^{86}$Kr/4; simultaneous analysis of $\delta^{86}$Kr/$^{83}$Kr and $\delta^{86}$Kr/$^{84}$Kr shows the fractionation to be mass-dependent. Because the ice and atmospheric samples have comparable $\delta^{86}$Kr values we can exclude the gas extraction step as the origin of the fractionation. The results suggest a fractionation in the $^{81}$Kr/$^{83}$Kr ratio of around -5‰ (-0.5%), an order of magnitude smaller than the 5-7 pMKr analytical uncertainty in the ATTA technique.

Summarizing, we contend that within the precision of our analyses the samples are free of gas loss and gas exchange due to surface fracturing, sampling or processing; isotopic fractionation during sample processing introduces errors that are well within the stated $^{81}$Kr dating uncertainty. Sampling the ice at depths ≥5m appears to be essential for obtaining unaltered air samples from Taylor Glacier; this depth horizon could be a function of the (micro-)climate and ice dynamics of the site, and may not be representative of other BIAs.

In the radiometric dating we make the assumption that the atmospheric $^{81}$Kr/Kr ratio has remained stable over the time period of interest. Changes in ocean temperature on the timescale of glacial cycles can modify the atmospheric Kr inventory through the dependence of gas solubility on temperature; this effect is on the order of 0.1%, and clearly insufficient to influence our results (48). The cosmogenic $^{81}$Kr production rate in the upper atmosphere is expected to vary in response to changing solar activity and geomagnetic field strength (49, 50). The atmospheric $^{81}$Kr abundance represents the integrated cosmogenic production rate over a time period that corresponds to the $^{81}$Kr lifetime ($t_{1/2}$/ln(2)=330 ka). Consequently, for all practical purposes the $^{81}$Kr abundance is insensitive to production variability on annual to millennial time scales related to solar cycles (51) and geomagnetic excursions such as the Laschamp event (52). Long term reconstructions of geomagnetic dipole strength show pronounced variations on multi-millennial time scales (53), as plotted in Fig 3a. To estimate the impact on atmospheric $^{81}Kr$ we converted the geomagnetic variations to cosmogenic nuclide production rates using the method of Wagner et al. (54), where we assume that $^{81}Kr$ production scales in the same way as $^{10}Be$ and $^{36}Cl$ production (the rationale behind this assumption is that $^{81}Kr$ production is dominated by spallation as is $^{10}Be$ and $^{36}Cl$ production, whereas $^{14}C$ production is the result of thermal neutron capture). Estimated past relative $^{81}Kr$ production rates and the resulting $^{81}Kr/Kr$ abundances are shown in Fig. 3b. Since the Brunhes–Matuyama reversal (780 ka ago) the geomagnetic field has been relatively strong, leading to increased cosmic ray shielding and reduced radionuclide production. Our estimate suggests that during the last 1.5 Ma atmospheric $^{81}Kr/Kr$ ratios have remained between 0.96 and 1.1 times the long-term average. The assumption of constant $^{81}Kr/Kr$ ratios introduces an error, the magnitude of which is shown in Fig. 3c. For the last 350 ka (which includes this study) this error is below 3ka, and well within analytical uncertainty. No correction was therefore applied to the $^{81}Kr$ ages. For the entire 1500 ka time interval of interest to dating studies, the estimated error remains below 4% of absolute age. In principle, if independently dated old ice is available, such as at the Mount Moulton BIA (13), $^{81}Kr$ can be used as a tracer of past cosmic ray variability. Such a $^{81}Kr$-based reconstruction would be insensitive to past changes in atmospheric transport or biogeochemistry, which is not the case for $^{10}Be$ and $^{14}C$, respectively (50).

Our result shows that $^{81}Kr$ radiometric dating of ancient ice is both feasible and accurate within the specified analytical uncertainty. In the 50 ka - 1.2 Ma age range our method is currently more accurate than other absolute dating tools for ice, such as the $^{40}Ar$ method (uncertainty ≥180 ka, (19)) or recoil U-series dating (uncertainty 16-300 ka (14)). $^{81}Kr$ dating therefore has great potential for the dating of BIAs, thereby enhancing their scientific value as archives of easily accessible old ice. Given the current accuracy of the ATTA method, a 10 μL Kr sample (around 100 kg of ice) would allow dating 1 Ma old ice within 100 ka age uncertainty, and 1.5 Ma old ice within 300 ka (32). For ice older than 1.2 Ma (around 5 $^{81}Kr$ half-lives) the $^{40}Ar$ method becomes the more accurate absolute dating technique. Kr sample size requirements for the ATTA method have decreased by almost 4 orders of magnitude since the first experimental realization (33), and currently a minimum of 40 kg of ice is required for a single analysis. If technological advances can further reduce sample requirements in the future, $^{81}Kr$ dating can be applied to regular ice core samples as well. It will be particularly helpful with traditionally difficult dating problems, such as basal ice.

Our study reveals that the Taylor Glacier BIA contains large quantities of ice from both the penultimate deglaciation and the previous interglacial period (MIS 5e, 130-115 ka BP) that can be sampled near the glacier surface. This is of interest to paleoclimatic reconstructions that require large ice samples, such as isotopic measurements of atmospheric trace gases (55, 56). The deglaciation is accompanied by a global reorganization of biogeochemical cycles, as evidenced by (abrupt) changes in the atmospheric abundance of trace gases such as $CH_4$, $CO_2$ and $N_2O$. Detailed records of the isotopic composition of these gases can help better constrain changes in their global budgets. The valuable MIS 5e ice can be sampled at Taylor Glacier with a much smaller logistical footprint than would be required for a deep drilling campaign, as the ice outcrops at the surface and the site is within easy helicopter reach of McMurdo station. The presence of MIS 5e ice at Taylor Glacier furthermore has implications for attempts to reconstruct past ice sheet stability and ice flow in the larger Taylor Dome region (10).

**Methods**

Ice was sampled using a 24 cm diameter electromechanical ice drill without drilling fluid. All ice sampling was done below 5 m depth, either as two 5 m cores or a single 10m core. No fractures were observed in any of the samples. Except for the 12.2 ka sample, all samples were transported to the field camp on a custom made sled behind a snowmobile. Samples were cleaned with an electropolished chisel, and placed in an on-site 670 L aluminum vacuum-melter tank (57). The tank was evacuated for at least 1.5 h, after which the sample was melted using propane burners underneath the tank (~3 h). Using a bubbler manifold the air was recirculated for 0.5 h through the meltwater to equilibrate the gases between the water and the headspace (58). Next, the headspace air was transferred into 35 L electropolished stainless steel sample flasks (transfer time 0.5 h) for shipment to the laboratory. Air extraction was completed within 12 h of drilling. To prevent the ice cores from warming up, drilling and sample handling were done during the coldest hours of the night when the sun dips below the Kukri Hills.

The Kr and Ar were separated from sample air at the University of Bern using molecular sieve absorption and Gas Chromatography (59). Duplicate $^{81}$Kr/Kr and $^{85}$Kr/Kr analysis was performed using a single atom counting technique (Atom Trap Trace Analysis) in the Laboratory for Radiokrypton Dating, Argonne National Laboratory (32-34), using 10 $\mu$L pure Kr per measurement. In the apparatus, atoms of a targeted isotope ($^{81}$Kr, $^{85}$Kr, or the control isotope $^{83}$Kr) are captured by resonant laser light into an atom trap, and counted by observing the fluorescence of the trapped atoms. For quality control in the analysis of environmental samples, the instrument is calibrated with a standard modern sample both before and after the analysis of every group of 2 – 3 environmental samples. $^{39}$Ar activity was measured in Bern using low-level decay counting (60). Kr stable isotope measurements were performed at Scripps Institution of Oceanography using traditional dual inlet Isotope Ratio Mass Spectrometry (IRMS). $CH_4$ was measured at Oregon State University using a melt-refreeze air extraction, followed by gas chromatography with a flame ionization detector (61). $\delta^{18}O_{atm}$ was analyzed at Scripps Institution of Oceanography using a melt-refreeze air extraction, followed by dual inlet IRMS; results are normalized to La Jolla air, and routine analytical corrections are applied (62, 63).

**Acknowledgements**

We want to thank the Polar Geospatial Center for their help with satellite image processing; X. Faïn, L. Mitchell, H. Schaefer, A. Schilt, T. Bauska, McMurdo Operations, the Berg Field Center (Kate, Martha and Bija) and helicopter crews for support in the field; Jon Edwards for his help with air content calculations; Matthew Puretz for krypton stable isotope mass spectrometry. The Center for Ice and Climate in Copenhagen, Denmark, hosted a seminar that led to the inception of the study. Constructive comments by two anonymous reviewers helped improve the manuscript. This work was funded through NSF OPP Grants 0838936 (to E.J.B.) and 0839031 (to J.P.S.); C.B. is supported by the NOAA Climate and Global Change Fellowship Program, Administered by the University Corporation for Atmospheric Research; W.J., Z.T.L., P.M., and the Laboratory for Radiokrypton Dating at Argonne are supported by DOE, Office of Nuclear Physics, under contract DE-AC02-06CH11357. Development of the ATTA-3 instrument was supported in part by NSF EAR-0651161.

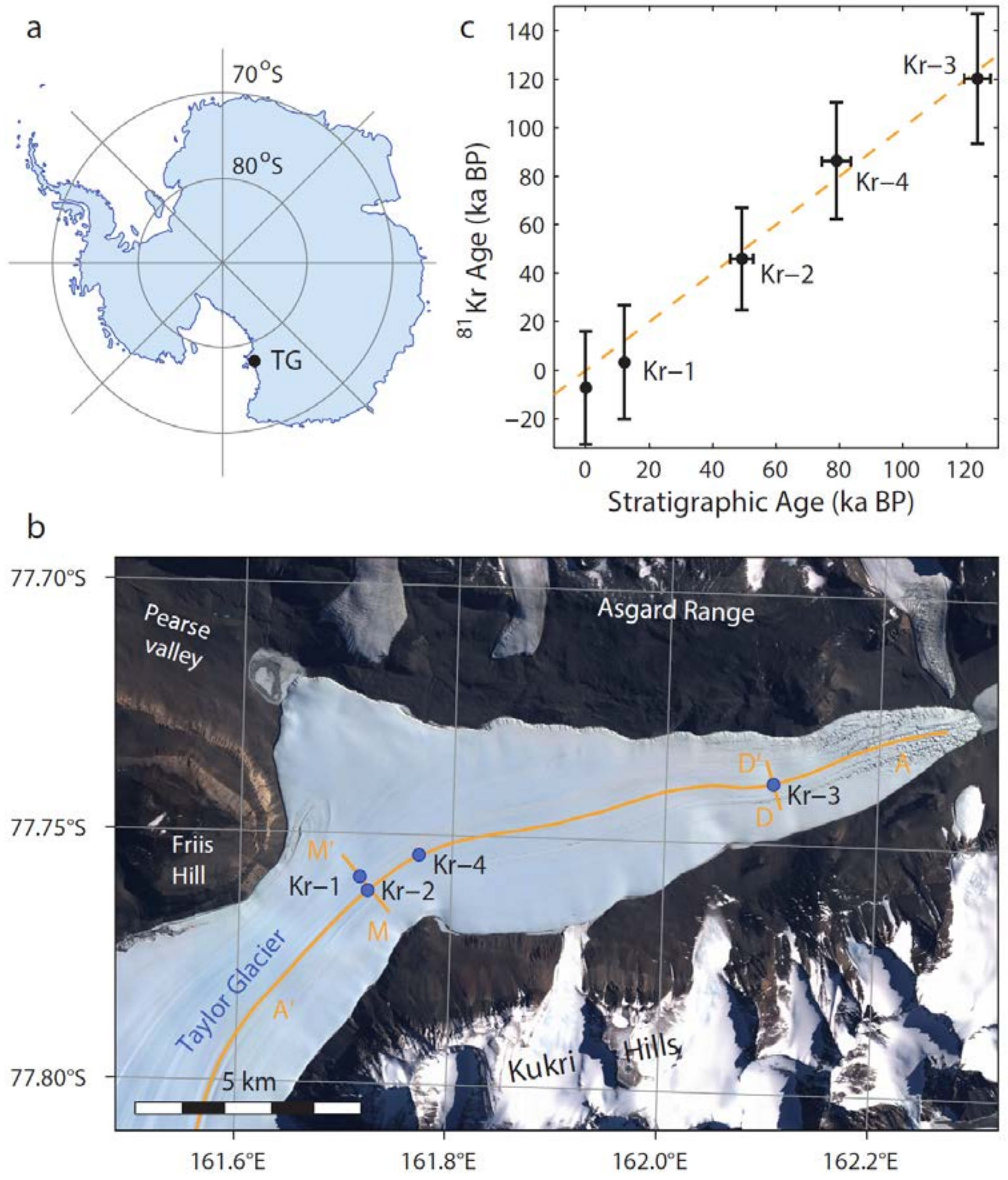

**Figure 1** Radiometric $^{81}$Kr dating at the Taylor Glacier BIA, Antarctica. a) Location of Taylor Glacier on map of Antarctica with the Greenwich meridian upwards. b) Satellite imagery of Taylor Glacier. Stratigraphically dated profiles indicated with A-A' (along flow), M-M' (main transect) and D-D' (downstream transect); length of perpendicular profiles (M and D) is not to scale. $^{81}$Kr sampling locations are indicated as blue dots. Image courtesy of DigitalGlobe, Inc. c) Comparison of $^{81}$Kr radiometric ages to independently derived stratigraphic ages, in thousands of years before 1950 C.E. (ka BP).

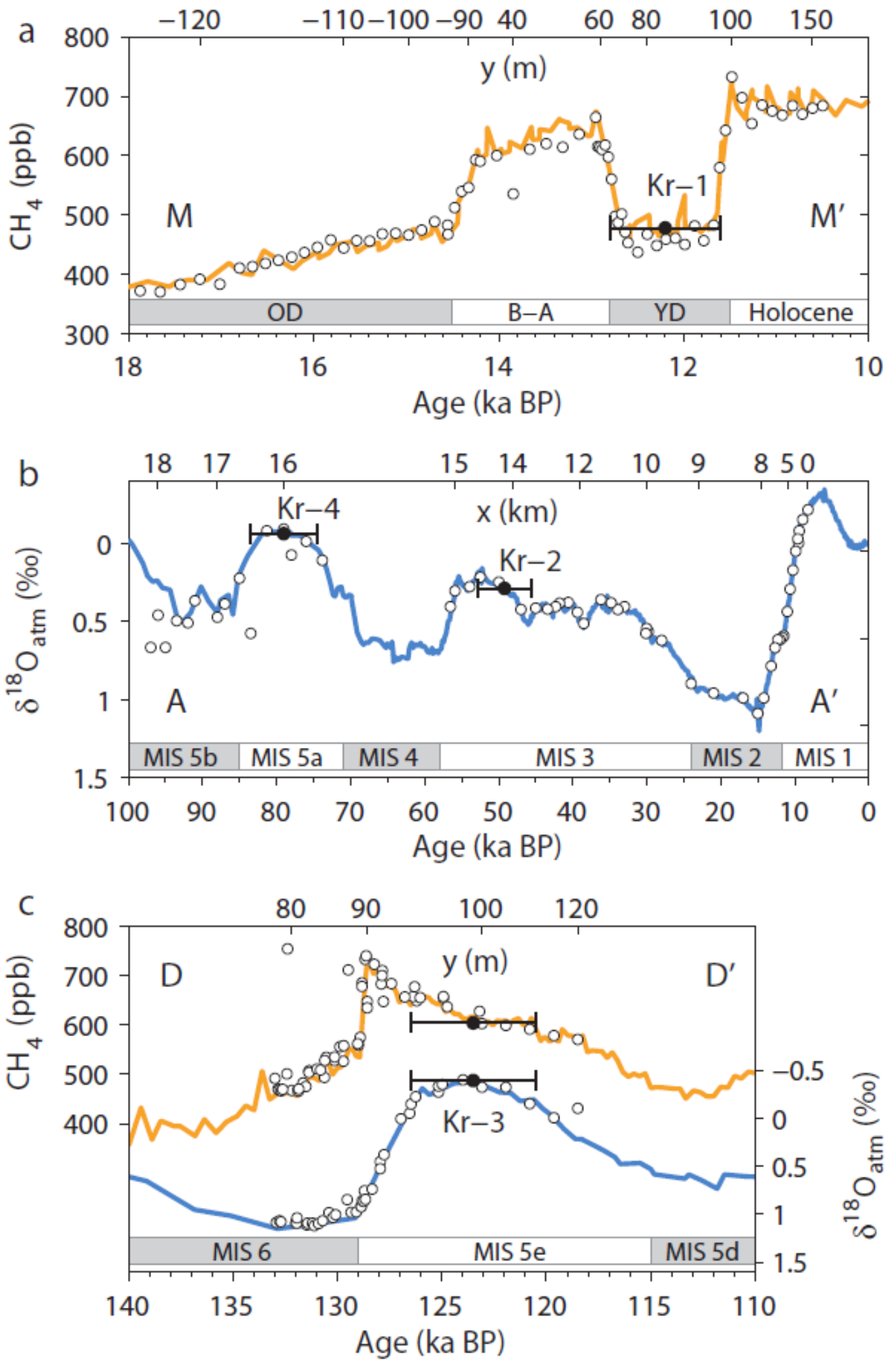

**Figure 2** Stratigraphic dating of Kr samples. Abbreviations are Oldest Dryas (OD), Bølling-Allerød (B-A) and Younger Dryas (YD). Measurements along the stratigraphically dated profiles (white dots) with Kr samples (black with age uncertainty). a) Sample Kr-1, located on the Main transect (see figure 1). $CH_4$ data from the EDML core, Antarctica (45). b) Samples Kr-2 and 4, located on the along-flow profile. $\delta^{18}O_{atm}$ data from the Siple Dome core, Antarctica (62). c) Sample Kr-3, located on the Downstream transect. $\delta^{18}O_{atm}$ ($\delta^{18}O$ of atmospheric $O_2$) data from EDML (64) and $CH_4$ data from EDC (65). The apparent phase shift between the $CH_4$ and $\delta^{18}O_{atm}$ rise across the penultimate deglaciation is mainly a function of the large difference in the atmospheric lifetime of $CH_4$ and $O_2$ (62). Taylor Glacier transect positions have been corrected for the isochrone dip angle (Supporting Information). Taylor Glacier $CH_4$ and $\delta^{18}O_{atm}$ data are not shown for the along-flow and main transects, respectively, as they do not provide a strong age constraint.

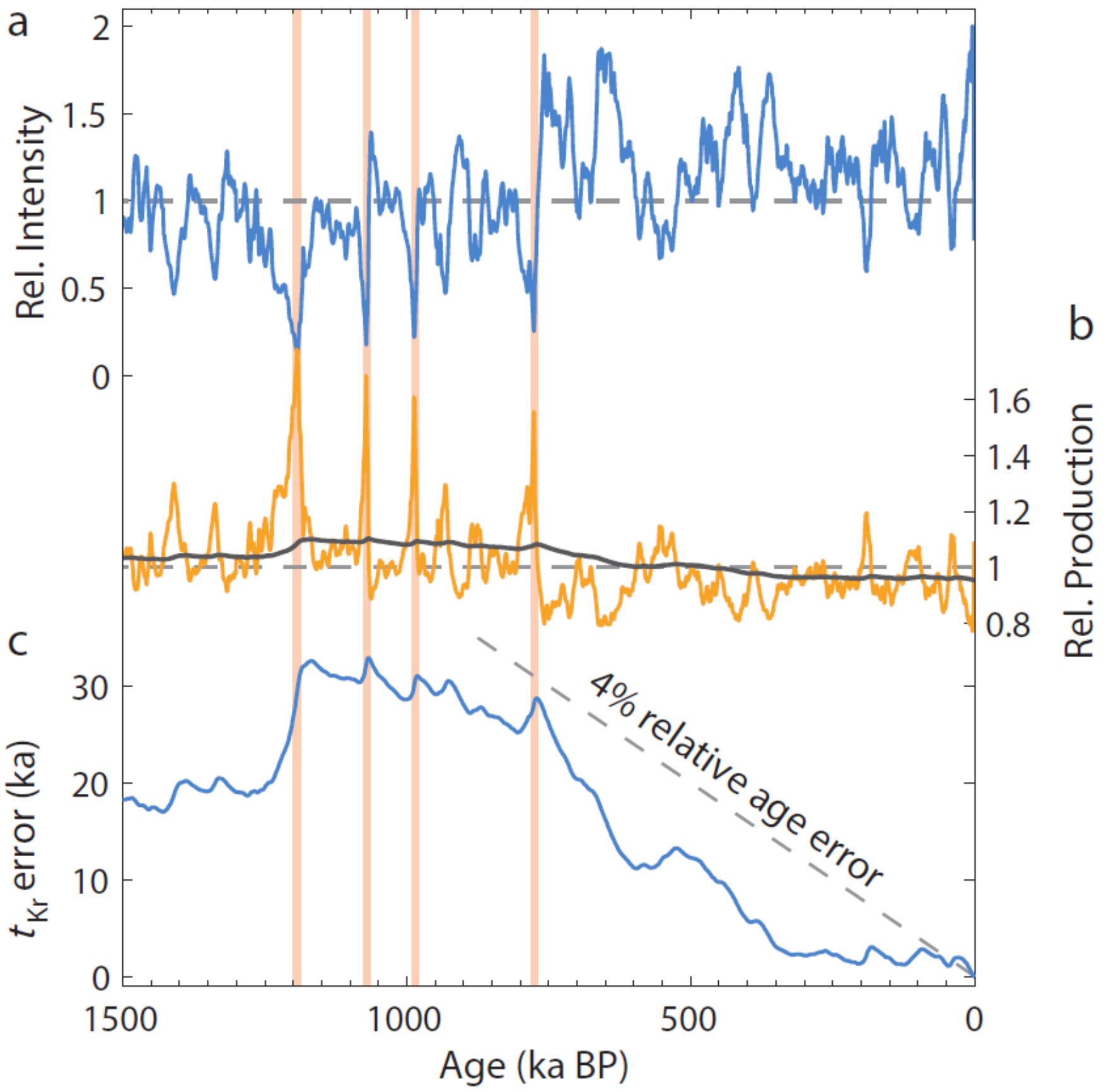

**Figure 3** Stability of atmospheric $^{81}$Kr. a) Relative paleointensity of the geomagnetic field (53). Magnetic reversals are indicated by vertical lines. b) Relative spallogenic production rate (orange) with relative $^{81}$Kr abundance (black). The $^{81}$Kr abundance is calculated through a convolution of the production rate with the atmospheric $^{81}$Kr residence (1/330·exp[-$t$/330], with $t$ the time in ka). c) Estimated error in $^{81}$Kr radiometric age when assuming stable atmospheric $^{81}$Kr/Kr; positive values indicate an underestimated $t_{Kr}$.

**Table 1.** Overview of Radiokrypton samples

| sample | Strat. age (ka BP) | $^{81}$Kr (pMKr) | $^{81}$Kr age† (ka BP) | $^{85}$Kr (dpm mL$^{-1}$) | $^{39}$Ar (pMAr) | $\delta^{86}$Kr/4 (‰) | air content (mL kg$^{-1}$) | $CH_4$ (ppb) | $\delta^{18}O_{atm}$ (‰) |
|---|---|---|---|---|---|---|---|---|---|
| Kr-1 | 12.2 ± 0.5 | 101 ± 7<br>97 ± 7 | 3 ± 23 | < 0.99<br>< 1.15 | 28 ± 13 | 3.51 | 110.8 | 477.8 | - |
| Kr-2 | 49 ± 3.6 | 94 ± 6<br>80 ± 5 | 46 ± 21 | < 0.89<br>< 1.10 | < 15 | 2.26 | 104.4 | 521.6 | 0.286 |
| Kr-3 | 123.5 ± 4.2 | 73 ± 6<br>66 ± 5 | 120 ± 26 | < 0.82<br>< 1.31 | < 30 | - | 107.0 | 604.3 | -0.390 |
| Kr-4 | 79 ± 4.5 | 76 ± 5<br>78 ± 6 | 86 ± 24 | < 0.74<br>< 1.33 | < 15 | 2.41 | 104.3 | 484.8 | -0.065 |
| Atm. | 0 | 107 ± 7<br>97 ± 7 | -7 ± 23 | 72.1 ± 3.8<br>70.9 ± 3.7 | - | 2.38 | - | 1755.0 | - |

†) duplicate $^{81}$Kr analyses are averaged to calculate the listed $^{81}$Kr radiometric